\documentclass[reqno]{article}

\usepackage{graphicx}
\usepackage{amsmath,amsthm,amssymb}

\chardef\bslash=`\\ 

\theoremstyle{definition}

\theoremstyle{remark}

\newcommand{\eval}[2][\right]{\relax
  \ifx#1\right\relax \left.\fi#2#1\rvert}

\begin{document}
\title{\bf{On blowup for semilinear wave equations\\ with a focusing nonlinearity}}

\author{Piotr Bizo\'n\footnotemark[1]{}{},\;
  Tadeusz Chmaj\footnotemark[2]{}\;,  and Zbis\l aw
  Tabor\footnotemark[3]{}\\{}\\
  \footnotemark[1]{} \small{\textit{Institute of Physics,
   Jagellonian University, Krak\'ow, Poland}}\\
   \footnotemark[2]{} \small{\textit{H.Niewodniczanski Institute of Nuclear
   Physics,
Polish Academy of Sciences, Krak\'ow,
    Poland}}\\ \footnotemark[3]{} \small{\textit{Department of Biophysics,
    Collegium Medicum,  Jagellonian University, Krak\'ow,
    Poland}}}
%
\maketitle
\begin{abstract}
\noindent In this paper we report on numerical studies of
formation of singularities  for  the semilinear wave equations
with a focusing power nonlinearity $u_{tt} - \Delta u = u^{p}$ in
three space dimensions. We show that for generic large initial
data that lead to singularities, the spatial pattern of blowup can
be described in terms of linearized perturbations about the
fundamental self-similar (homogeneous in space) solution. We
consider also non-generic initial data which are fine-tuned to the
threshold for blowup and identify critical solutions that separate
blowup from dispersal for some values of the exponent $p$.
\end{abstract}
\section{Introduction}
One of the most interesting features of many nonlinear evolution
equations is the spontaneous onset of singularities in solutions
starting from  smooth initial data. Such a phenomenon, usually
called "blowup", has been a subject of intensive studies in many
fields ranging from fluid dynamics to general relativity. Given a
nonlinear evolution equation, the key question is whether or not
the blowup can occur for some initial data.  Once the existence of
blowup is established for a particular equation, many further
questions come up, such as: When and where does the blowup occur?
What is the character of blowup and is it universal? What happens
at the threshold of blowup?

In this paper we consider these questions for the simplest
nonlinear generalization of the free wave equation: the semilinear
wave equation with the power nonlinearity
\begin{equation}\label{eqgen}
u_{tt} - \Delta u = u^{p}, \qquad  u = u(t,x), \quad x \in R^3.
\end{equation}
where $p>1$ is an odd integer. The sign of the nonlinear term
corresponds to focusing, that is it tends to magnify the amplitude
of the wave. If $u$ is small this term is negligible and  the
evolution is essentially linear (actually one has scattering for
$t\rightarrow \infty$). However, if $u$ is large the dispersive
effect of the linear wave operator may be overcome by the focusing
effect of the nonlinearity and a singularity can form. In fact, it
is known that if the energy
\begin{equation}\label{energy}
    E[u] = \int_{R^3}
 \left( u_t^2 + (\nabla u)^2 - \frac{1}{p+1} u^{p+1} \right) d^3 x
\end{equation}
is negative, then a singularity must form in a finite time
\cite{levine}. This theorem  says only that the solution cannot be
continued beyond certain time but it gives no information on how
the solution looks like as it approaches the blowup time. Probably
the best way to learn about the character of singularities is to
look at explicit singular solutions. For equation (\ref{eqgen}) it
is easy to see that
\begin{equation}\label{explicit}
u_0 = \frac{a}{(T-t)^{\alpha}}, \qquad
a=\left[\frac{2(p+1)}{(p-1)^2}\right]^{\frac{1}{p-1}}, \quad
\alpha=\frac{2}{p-1},  \quad T>0,
\end{equation}
is the exact solution which blows up as $t \rightarrow T$. This
solution is obtained by neglecting the laplacian in (\ref{eqgen})
and solving the corresponding ordinary differential equation
$u_{tt}=u^p$. The question is how typical this explicit singular
behaviour is. There are several ways to approach this problem. On
the analytical side there are  Fuchsian methods developed by
Kichenassamy \cite{kichen} which  allow to construct open sets of
initial data which blow up on a prescribed spacelike hypersurface
with the leading order asymptotic behaviour given by the solution
$u_0$ \cite{kichen2}. On the heuristic side there are numerical
and perturbative methods which, albeit non-rigorous, allow to gain
a more detailed information about the character of blowup. In this
paper we take the latter approach.

  Our main goal was to show that
  the solution $u_0$ determines the
  leading order asympotics of blowup for generic large initial
  data and the spatial pattern of convergence to this solution can
  be described in terms of the least damped eigenmodes of the linearized
  perturbations about $u_0$. We did this
in the spherically symmetric case
\begin{equation}\label{eqspher}
u_{tt} -  u_{rr} - \frac{2}{r}u_r = u^{p}
\end{equation}
  for
 three representative values  of $p=3$, $5$, and $7$.
These values  corresponds to three different classes of
criticality of equation (\ref{eqgen}). To see this, notice that
equation (\ref{eqgen}) has the scaling symmetry: if $u(t,x)$ is
the solution, so is
\begin{equation}\label{scale}
u_L(t,x) = L^{\alpha} u(t/L,x/L), \qquad \alpha = \frac{2}{1-p}.
\end{equation}
Under this transformation the energy  scales as
\begin{equation}\label{energy_scaling}
E[u_L] = L^{\beta} E[u], \qquad \beta = \frac{p-5}{p-1},
\end{equation}
 hence
 equation (\ref{eqgen}) is subcritical  for $p=3$ ($\beta<0$),
critical  for $p=5$ ($\beta=0$) and supercritical  for $p>5$
($\beta>0$). Since the energy (\ref{energy}) is not positive
definite, this distinction is not very important as far as the
generic character of blowup is concerned, however, as we shall
show below, it is relevant for understanding the behaviour of
solutions at the threshold for blowup.

The paper is organized as follows. In section 2 we discuss
self-similar solutions of equation (\ref{eqspher}) and analyze
their stability. In section 3 we present the results of numerical
simulations of blowup and demonstrate the universality of the
blowup profile. Finally, in section 4 we  discuss the behaviour of
solutions at the threshold for blowup.
\section{Self-similar solutions}
In order to set the stage for the discussion of singularity
formation we first discuss self-similar solutions of equation
(\ref{eqspher}). As we shall see below these solutions play an
important role in the process of blowup. By definition,
self-similar are invariant under rescaling (\ref{scale}), hence in
the spherically symmetric case they have the form
\begin{equation}\label{ssansatz}
u(t,r)= (T-t)^{-\alpha}\; U(\rho), \qquad \alpha = \frac{2}{p-1},
\quad \rho=\frac{r}{T-t},
\end{equation}
where $T$ is a positive constant, clearly allowed by time
translation invariance.  Note that each self-similar solution, if
it is regular for $t<T$, provides an explicit example of a
singularity developing at $r=0$ in finite time $T$ from
nonsingular initial data - for this reason we shall refer to $T$
as the blowup time. Substituting the ansatz (\ref{ssansatz}) into
equation (\ref{eqspher}) one gets the ordinary differential
equation for the similarity profile $U(\rho)$
\begin{equation}\label{eqss}
(1-\rho^2)U'' + \left( \frac{2}{\rho} - (2 + 2 \alpha) \rho
\right) U' - \alpha (\alpha + 1) U + U^p =0.
\end{equation}
We consider this equation inside the past light cone of the blowup
point $(t=T, r=0)$, that is in the interval $0 \leq \rho \leq 1$.
It is easy to see that for any $p$ equation (\ref{eqss}) has  the
constant solution
\begin{equation}\label{ss_const}
U_0(\rho)=\left[\frac{2(p+1)}{(p-1)^2}\right]^{\frac{1}{p-1}}.
\end{equation}
Of course, this solution corresponds exactly to the solution $u_0$
of the original equation (\ref{eqspher}). It turns out that
besides this trivial solution, for some values of $p$ there exist
also nontrivial  profiles. The existence of such solutions can be
proved by the shooting technique which goes as follows. One first
shows that at the both endpoints of the interval $0 \leq \rho \leq
1$ there exist one-parameter families of local analytic solutions
which behave, respectively, as
\begin{equation}\label{local0}
U(\rho) \sim c +  \frac{1}{3} \left( \frac{p+1}{(p-1)^2} c - \frac{1}{2} c^p \right) \rho^2
\quad \text{for} \quad \rho \rightarrow 0,
\end{equation}
and
\begin{equation}\label{local1}
U(\rho) \sim b + \frac{1}{2} \left( \frac{1}{2}  (p-1)\, b^p - \frac{p+1}{p-1}\, b \right) (\rho -1)
\quad \text{for} \quad \rho \rightarrow 1,
\end{equation}
where $b$ and $c$ are free parameters. Having that, the strategy
for finding solutions which are regular in the interval $0 \leq
\rho \leq 1$
 is simple: one shoots the solution satisfying the initial condition (\ref{local1}) at $\rho=1$
towards the center and adjusts the parameter $b$ so that
$U'(0)=0$. Applying this technique Bizo\'n and Maison \cite{bm}
proved existence for a countable set of parameters $b_n$
($n=0,1,\ldots$) which determine analytic similarity profiles
$U_n$ for $p=3$ and all odd $p\geq 7$. The first few similarity
profiles $U_n$ for $p=3$ and $p=7$ are shown in figures 1 and 2.

  The behaviour of solutions $U_n(\rho)$ outside the past light
  cone, that is for $\rho>1$, depends on $p$. One can show (see
  \cite{bm}) that for $p=3$ all $n>0$ solutions become singular
outside the past light cone, namely
\begin{equation}
U(\rho) \sim \frac{d}{\rho_0 - \rho} \quad \text{for some} \quad
\rho_0 > 1.
\end{equation}
 In contrast, for $p=7,9,\ldots$ all solutions $U_n$ remain
regular
 outside the past light cone.
 \pagebreak
 \begin{figure}[h!]
\centering
\includegraphics[width=0.6\textwidth]{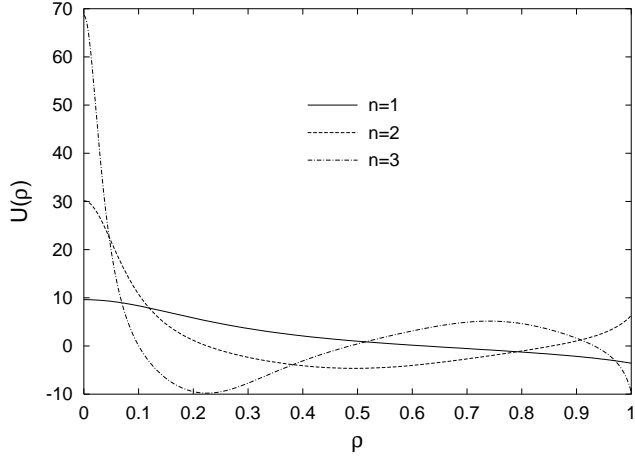}
\vskip 0.15cm
 \caption{\small{The first three similarity profiles $U_n(\rho)$ for $p=3$. The index $n$ counts
the number of zeros of $U_n(\rho)$ in the interval $0 \leq \rho
\leq 1$. When continued beyond the past light cone, these
solutions become singular at some $\rho>1$.}}
\end{figure}
\begin{figure}[h!]
\centering
\includegraphics[width=0.6\textwidth]{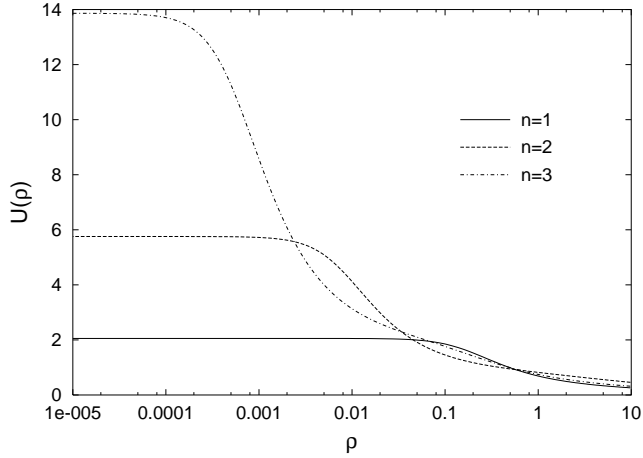}
\vskip 0.15cm
 \caption{\small{The first three similarity profiles $U_n(\rho)$ for $p=7$.
 In contrast to $p=3$, here all profiles are monotone and have no
 zeros. For $\rho
\rightarrow \infty$ they decay as $U(\rho) \sim \rho^{-1/3}$.}}
\end{figure}

\noindent For $p=5$ there are no nontrivial self-similar solutions
- this can be showed as follows. Consider the function
\begin{equation}\label{q}
    Q(\rho) = \frac{1}{2} (1-\rho^2) \rho^3 U'^2 + \frac{1}{2}
    \rho^2 (1-\rho^2) U U' +
    \left[\frac{3 (5-p)}{4 (p-1)}-\frac{2}{(p-1)^2}\right] \rho^3 U^2 +
    \frac{1}{p+1} \rho^3 U^{p+1}.
\end{equation}
This function was  introduced by Kavian and Weissler \cite{kw} in
their study of equation (\ref{eqspher}). They showed that
$Q'(\rho)=0$ for $p=5$, hence in this case $Q$ is the first
integral of equation (\ref{eqss}). Since $Q(0)=0$, it follows that
$Q(1)=0$, from which one gets $b=U(1)=(3/4)^{1/4}$. This coincides
with $U_0$  so by uniqueness we conclude that $U_0$ is the only
similarity profile.

In order to determine the role of self-similar solutions in
dynamics it is essential to analyze their stability. To this end
we define the
 slow time $\tau = -\ln(T-t)$ and rewrite equation (\ref{eqspher}) in terms
 of the new variable $U(\tau, \rho)$ defined by
\begin{equation}
u(t,r)=e^{\alpha \tau} U(\tau,\rho), \qquad \alpha=\frac{2}{p-1}.
\end{equation}
We get
\begin{equation}\label{css}
U_{\tau\tau} + (1+2 \alpha) U_{\tau} + 2 \rho U_{\tau\rho} =
(1-\rho^2) U_{\rho\rho}  + \left( \frac{2}{\rho} - (2 + 2 \alpha)
\rho \right) U_{\rho} - \alpha (\alpha + 1) U + U^p.
\end{equation}
The advantage of this formulation is that  self-similar solutions
of equation (\ref{eqspher}) correspond now to $\tau$-independent
solutions  of equation (\ref{css}) while the asymptotics of blowup
corresponds to the behaviour at $\tau \rightarrow \infty$.
 In order to determine the linear stability of solutions $U_n(\rho)$
 we seek solutions of (\ref{css}) in the form
 $U(\tau,\rho)=U_n(\rho) + e^{\lambda \tau} \xi(\rho)$.
After linearization we get the quadratic eigenvalue problem
\begin{equation}\label{eigen_fn}
(1-\rho^2) \xi'' + \left( \frac{2}{\rho} - 2 (1+\alpha) \rho
\right) \xi' - 2 \rho \lambda \xi'
   + \left[ p U_n^{p-1} - \alpha (\alpha+1) - (1+2 \alpha) \lambda - \lambda^2 \right] \xi =0.
\end{equation}
Let us consider first the stability of the constant solution
$U_0$.  In this case equation (\ref{eigen_fn}) becomes
\begin{equation}\label{eigen_f0}
(1-\rho^2) \xi'' + \left[ \frac{2}{\rho} - 2 \, \left( \frac{p+1}{p-1} + \lambda \right) \, \rho \right] \xi' +
      \left[ \frac{2(p+1)}{p-1} - \frac{p+3}{p-1} \, \lambda - \lambda^2 \right] \xi =0.
\end{equation}
Near $\rho=0$ the admissible solution has the formal power series
expansion
\begin{equation}\label{series_exp}
\xi(\rho)=\sum_{k=0} a_k \rho^{2k}
\end{equation}
with the coefficients satisfying the recurrence relation
\begin{equation}
a_{k+1}={\lambda^2 + \left( 4k + \frac{p+3}{p-1} \right) \lambda +
2k \left( 2k + \frac{p+3}{p-1} \right)  -2 \, \frac{p+1}{p-1}\over
2 (k+1) (2k+3) } \,a_k .
\end{equation}
Since $a_{k+1}/a_k \rightarrow 1$ as $k \rightarrow \infty$, the series (\ref{series_exp}) diverges for $\rho > 1$.
In order to pass smoothly through the point $\rho=1$ (so that the eigenfunctions are well-behaved also outside
the past light cone) we impose the condition that the series truncates at the $k$-th term
\begin{equation}
\lambda^2 + \left( 4k + \frac{p+3}{p-1} \right) \lambda + 2k \left( 2k + \frac{p+3}{p-1} \right)  - 2 \,\frac{p+1}{p-1} =0.
\end{equation}
This yields two infinite sequences of pairs of real eigenvalues
\begin{equation}\label{eigen_val}
\lambda_k = 1 - 2 k, \quad
\overline{\lambda}_k=-\frac{2(p+1)}{p-1} -2k, \qquad k=0, 1,
\ldots .
\end{equation}
There is exactly one positive eigenvalue $\lambda_0 =1$. It
corresponds to the gauge mode which is due to the freedom of
choosing the blow-up time $T$. All the remaining eigenvalues are
negative hence for any $p$ the solution $U_0$  is linearly stable.
This suggests that it can appear as an attractor in generic
evolution.

Since we do not know the solutions $U_n$ with $n>0$ in closed
form, their spectrum of linear perturbations can be computed only
numerically. Our numerical calculations indicate that the solution
$U_n$ has $n$ unstable modes (apart from the spurious unstable
mode corresponding to the change of blowup time). For this reason
the solutions with $n>0$ are not expected to appear in generic
evolution. However, as we shall see below, the solution $U_1$ with
one unstable mode  appears as the codimension-one attractor in the
evolution of specially prepared initial data.
\section{Blowup profile}
Having learned about the stability of the solution $u_0$ we are
now prepared to interpret the results of numerical simulations.
The main goal of these simulations was to determine the
asymptotics of blowup. We solved equation (\ref{eqspher}) using
the method of lines which was fourth order accurate in space and
time. We found that for sufficiently large initial data the
amplitude $u(t,r)$ becomes unbounded  in a finite time $T$ for
some $r=r_S$. More precisely, we have
\begin{equation}\label{blow}
    \lim_{t\rightarrow T} (T-t)^{\alpha} u(t,r_S) = a =
    \left[\frac{2(p+1)}{(p-1)^2}\right]^{\frac{1}{p-1}},
\end{equation}
which confirms the expectation that the solution $u_0$ determines
the leading order asymptotics of blowup. In this section we wish
to show that if the blowup point is at the center, i.e. $r_S=0$,
then the spatial pattern of the developing singularity can be
described in terms of the least damped eigenmodes about $u_0$.

Using the results of linear stability analysis we can represent
the asymptotic approach to $U_0$ for $\tau \rightarrow \infty$
(i.e. $t \rightarrow T$)  by the formula
\begin{equation}\label{eigen_exp}
U(\tau,\rho)=U_0 + \sum_{k=1}c_k e^{\lambda_k \tau} \xi_k(\rho) +
             \sum_{k=0} \overline{c}_k e^{\overline{\lambda}_k \tau} \overline{\xi}_k(\rho),
\end{equation}
where $\xi_k(\rho), \, \overline{\xi}_k(\rho)$ are the eigenmodes
corresponding to the eigenvalues $\lambda_k, \,
\overline{\lambda}_k$, respectively and $c_k,\, \overline{c}_k$
are the expansion coefficients. Keeping the first two least damped
eigenmodes we obtain the following expansions in terms of original
variables (using the abbreviation $\delta = T-t$)
\begin{description}
\item[$p=3$]
\begin{equation}\label{expan_3}
\delta \, u(r,t)= \sqrt{2} + c_1 \delta \left( 1-\frac{r^2}{\delta^2} \right) + c_2 \delta^3 \left( 1-\frac{2 r^2}{3 \delta^2}
+ \frac{r^4}{5 \delta^4} \right) + O(\delta^4),
\end{equation}
\item[$p=5$]
\begin{equation}\label{expan_5}
\sqrt{\delta} \, u(t,r)={\left(\frac{3}{4}\right)}^{1/4} + c_1 \delta \left( 1-\frac{2 r^2}{3 \delta^2} \right) +
\delta^3 \left( \overline{c}_0 + c_2 \frac{r^2}{\delta^2} (1-\frac{r^2}{5 \delta^2}) \right) + O(\delta^5),
\end{equation}
\item[$p=7$]
\begin{equation}\label{expan_7}
\delta^{1/3}\, u(t,r)={\left(\frac{2}{3}\right)}^{1/3} + c_1
\delta \left( 1-\frac{5 r^2}{9 \delta^2} \right) + \overline{c}_0
\delta^{8/3} + O(\delta^3) .
\end{equation}
\end{description}
We claim that these formulae describe accurately the convergence
to the blowup profile inside the past light cone of the blowup
point $(t=T, r=0)$. The numerical evidence for this assertion is
summarized in figures 3 and 4 in the case $p=3$ (throughout this
section we use the case $p=3$ for illustration - analogous results
hold for $p=5$ and $7$).
\begin{figure}[h!]
\centering
\includegraphics[width=0.63\textwidth]{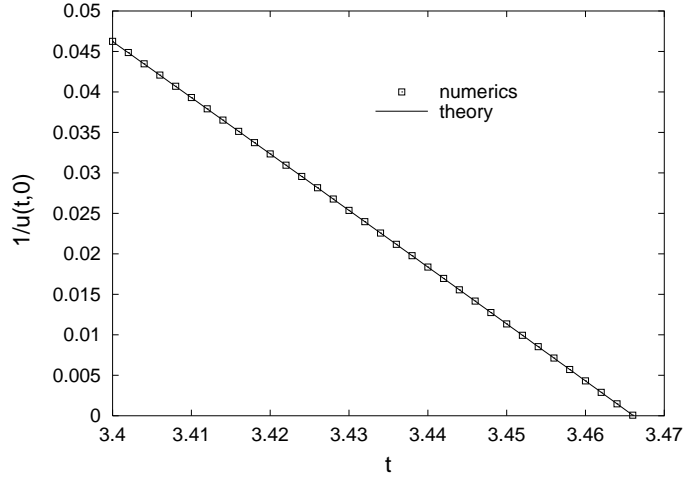}
\vskip 0.15cm
 \caption{\small{For $p=3$ we plot the function $1/u(t,0)$
for the solution that blows up at $r=0$ as $t\rightarrow T \approx
3.466$. The solid line shows the fit to the first order analytic
approximation $\delta/(\sqrt{2}+c_1 \delta)$.}}
\end{figure}
\begin{figure}[h!]
\centering
\includegraphics[width=0.63\textwidth]{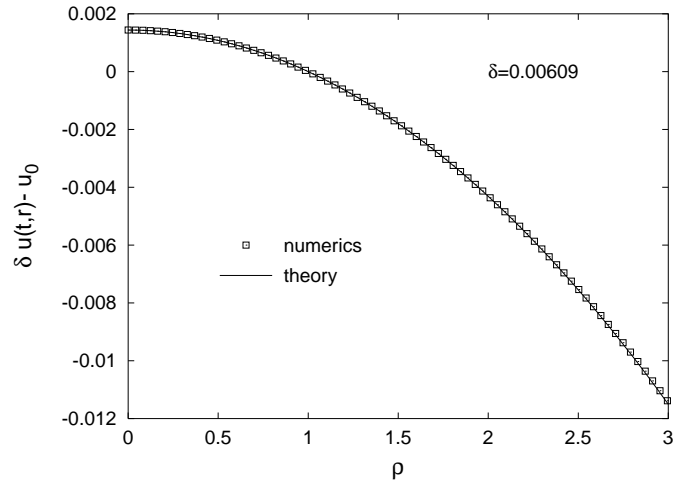}
\vskip 0.15cm
 \caption{\small{For the same numerical data as in figure 3 we plot the deviation of the rescaled solution
$\delta u(t,r/\delta)$ from the constant solution $U_0=\sqrt{2}$
at  time $\delta=6.09\times 10^{-3}$. The solid line shows the
least damped  eigenmode $c_1 (1-\rho^2) \delta$ with the same
coefficient $c_1$ as in figure 3.}}
\end{figure}
\begin{figure}[h!]
\centering
\includegraphics[width=0.61\textwidth]{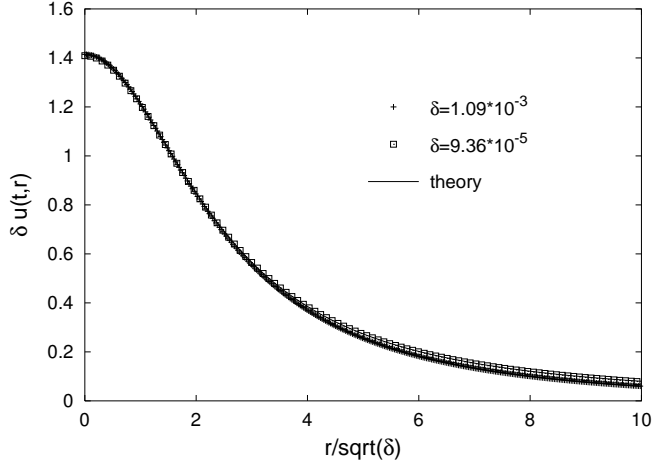}
\vskip 0.10cm
 \caption{\small{For $p=3$ the rescaled solution at two moments of time close to the blowup time
 is shown to collapse to  the analytic curve $F(z)=\sqrt{2}/(1+b
z^2)$ with $b=c_1/\sqrt{2}$.}}
\end{figure}

\noindent As shown in figure 4 the formula (\ref{expan_3})
accurately describes the blowup profile for large $\tau$ (i.e. $t$
close to $T$) not only inside the light cone but even slightly
beyond. However, the expansions (\ref{expan_3}-\ref{expan_7}) are
expected to break down when the linearization is no longer valid;
that is, if $r^2/\delta^2 \sim 1/\delta$. In this transition
region the leading order approximation for any $p$ reads
\begin{equation}\label{parabolic_0}
u(t,r) \simeq \frac{1}{\delta^{\alpha}} \left(a + c_1 d_{12}
\frac{r^2}{\delta} \right),
\end{equation}
with $\alpha$ and $a$  defined as in (\ref{explicit}) and $d_{12}$
equal to the coefficient of the quadratic term of the $\xi_1$
eigenfunction. This indicates the parabolic scaling
\begin{equation}\label{parabolic_1}
u(t,r) = \frac{1}{\delta^{\alpha}} F(z), \quad
z=\frac{r}{\sqrt{\delta}}.
\end{equation}
Substituting this ansatz into equation (\ref{eqspher}) and
dropping the laplacian (which becomes negligible as $\delta
\rightarrow 0$) we get the ordinary differential equation
\begin{equation}
z^2 F'' + (4 \alpha +3) z F' + 4 \alpha (\alpha +1) F - 4 F^p =0,
\end{equation}
which has a one-parameter family of regular solutions
\begin{equation}\label{parabolic_sol}
F(z)=\frac{a}{\left( 1+bz^2 \right)^\alpha}.
\end{equation}
Comparing (\ref{parabolic_0}) with (\ref{parabolic_1}) and
(\ref{parabolic_sol}) we get the matching condition
\begin{equation}
b=-\frac{d_{12}}{\alpha a} \,c_1,
\end{equation}
which e.g. for $p=3$ gives  $b=c_1/\sqrt{2}$. The numerical
confirmation of this prediction is shown in figure 5.

We remark  that the above result follows immediately from the
Fuchsian analysis which predicts the leading order asymptotics on
a spacelike blow-up curve $T(r)$ in the form
\begin{equation}\label{fuchs_form}
u(t,r)= \frac{a}{\left( T(r)-t \right)^\alpha}.
\end{equation}
The blowup time is defined as $T=\inf T(r)$. Assuming that this
infimum is attained at $r=0$, we have $T(r) \simeq T + b r^2$ for
some $b>0$. Inserting this into (\ref{fuchs_form}) we get
\begin{equation}
u(t,r)=\frac{1}{\delta^\alpha} \frac{A}{\left(1+b
\frac{r^2}{\delta}\right)^\alpha},
\end{equation}
which reproduces (\ref{parabolic_1}) and (\ref{parabolic_sol}).
\begin{figure}[hb]
\centering
\includegraphics[width=0.6\textwidth]{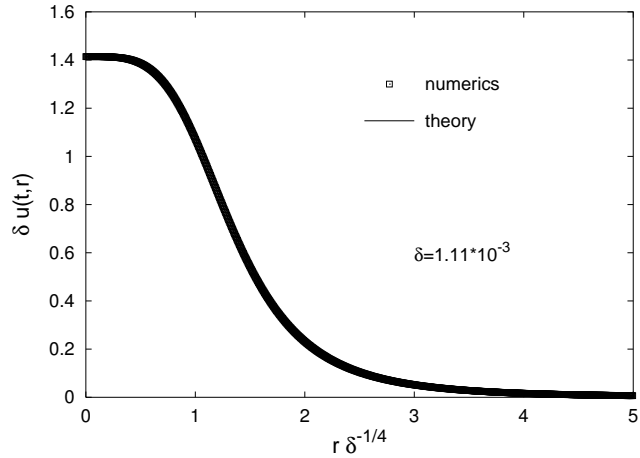}
\vskip 0.10cm
 \caption{\small{For $p=3$ we plot the rescaled solution approaching  the blowup time
for initial data fine-tuned to the borderline of blowup at $r_S=0$
and $r_S>0$.
 The solid line shows the fit to the analytic  prediction
$G(z)=\sqrt{2}/(1+d z^4)$}.}
\end{figure}

\noindent As the coefficient $b$  becomes negative, the first
blowup occurs at $r_S>0$. By fine-tuning  initial data to the
transition between the blowup at $r_S=0$ and the blowup at $r_S>0$
we set $b=0$ which means that the first eigenmode in the expansion
(\ref{eigen_exp}) is tuned away. For such codimension-one initial
data the formula (\ref{parabolic_0}) should be replaced by
\begin{equation}\label{quartic_0}
u(t,r) \simeq \frac{1}{\delta^{\alpha}} \left(a + c_2 d_{23}
\frac{r^4}{\delta} \right),
\end{equation}
where the coefficient $d_{23}$ is equal to the quartic term of the
$\xi_2$ eigenfuction. This gives  another scaling (see figure 6)
\begin{equation}\label{quartic_1}
u(t,r) = \frac{1}{\delta^{\alpha}} G(z), \qquad
z=\frac{r}{\delta^{1/4}},
\end{equation}
where
\begin{equation}\label{quartic_sol}
G(z)=\frac{a}{\left( 1+d z^4 \right)^\alpha}, \qquad d =
-\frac{d_{23}}{\alpha a} \,c_2.
\end{equation}
\section{Threshold for blowup}
Since  solutions of equation (\ref{eqspher}) disperse for small
initial data and blow up for large initial data, there arises a
natural question what happens in between. In the following the
boundary between initial data  that lead to dispersion and initial
data that lead to singularity formation  will be referred to as
the threshold for blowup. The determination of the threshold for
blowup and the corresponding dynamics  is of great interest in
physical models which predict formation of singularities, for
example in general relativity. This issue  can be studied
numerically as follows. Consider a one-parameter family of initial
data $\phi(p)$ such that the corresponding solutions exist
globally if the parameter $p$ is small or blow up if the parameter
$p$ is large. Then, along the curve $\phi(p)$ there must be a
critical value $p^*$ (or an interval $[p^*_{min},p^*_{max}]$)
which separates these two scenarios. Given two values $p$ small
and $p$ large, it is straightforward (in principle but not always
in practice) to find $p^*$ by bisection. Repeating this procedure
for different interpolating families of initial data one obtains a
set of critical data  which by construction belongs to the
threshold for blowup. Having that, one can look in more detail at
the evolution of critical data. The precisely critical data cannot
be prepared numerically but in practice it is sufficient to follow
the evolution of marginally critical data. Typically, one finds
that the evolution of such data has a universal (that is family
independent) transient phase during which the solution approaches
a kind of an intermediate attractor.
\begin{figure}[h!]
\centering
\includegraphics[width=0.7\textwidth]{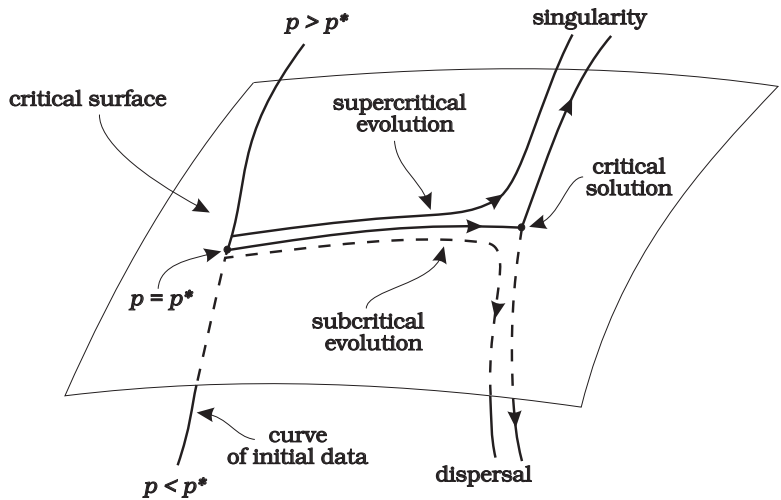}
\vskip 0.15cm
 \caption{\small{A schematic phase space picture of dynamics at the threshold for blowup}.}
\end{figure}

\noindent The heuristic explanation of this behaviour is sketched
in figure 7.
 According to this
picture the threshold for blowup is given by the codimension-one
stable manifold $W_S(u^*)$ of an intermediate attractor $u^*$,
called the critical solution. The critical initial data
corresponding to intersections of $W_S(u^*)$ with different
interpolating one-parameter families of initial data,
converge\footnote{It should be stressed that for conservative wave
equations, such as (\ref{eqgen}), the convergence (which is due to
radiation of energy to infinity) is always meant in the local
sense.} along $W_S(u^*)$ towards the critical solution. The
marginally critical data, by continuity, initially remain close to
$W_S(u^*)$ and approach $u^*$ for intermediate times but
eventually are repelled from its vicinity along the
one-dimensional unstable manifold. Within this picture the
universality of marginally critical dynamics in the intermediate
asymptotics follows immediately from the fact that the same
unstable mode dominates the evolution of all solutions. The nature
of the critical solution itself depends on a model - typically
$u^*$ is a static or a self-similar solution with exactly one
unstable mode.

To apply the numerical strategy outlined above we solved equation
(\ref{eqspher}) for various one-parameter families
  of initial conditions which
interpolate between small and large initial data. The results
described below do not depend on the particular choice of the
family -- for concreteness we present them for the  initial data
of the form
\begin{equation}\label{gauss4}
u(0,r)= A \, r^2 \, \text{exp} \left[ - \left( \frac{r-R}{\sigma}
\right)^4 \right], \quad u_t(0,r)=0,
\end{equation}
with adjustable amplitude $A$ and fixed parameters $\sigma$ and
$R$. Since the initial data are time symmetric, the initial
profile splits into ingoing and outgoing waves travelling with
approximately unit speed. Except for very large initial amplitudes
for which the singularity forms very fast, before the separation
into ingoing and outgoing wave occurs, the evolution of the
outgoing wave does not affect the singularity formation so we
shall ignore it. The behaviour of the ingoing wave depends on the
amplitude $A$. For large amplitudes we observe the formation of
singularity at some $r_S>0$ in finite time $T$.  As $A$ decreases,
the blowup point $r_S$  decreases also and reaches\footnote{The
behaviour of the function $r_S(A)$ depends on $p$. As $A
\rightarrow A_0$ from above, the function $r_S(A)$ decreases
continuously to zero  for $p=3$ but for $p>3$ it jumps from some
$r_S>0$ to $r_S=0$.} $r_S=0$ for some value $A_0$. As we keep
decreasing the amplitude below $A_0$ we eventually reach a
critical value $A^*$ below which solutions do not blowup up. The
asymptotic pattern of blowup described in the previous section
applies to solutions with amplitudes $A^*<A<A_0$. The character of
the threshold for blowup at $A^*$ depends on $p$ so we discuss
three values of $p$ separately. \vspace{-0.4cm}
\subsection*{$p=7:$} In this case we identify the critical solution
as the $n=1$ self-similar solution
\begin{equation}\label{crit7}
    u_1(t,r)=  (T-t)^{-\frac{1}{3}}\; U_1(\rho).
\end{equation}
The numerical evidence for the criticality of  solution $u_1$ is
presented in figure 8.
\begin{figure}[h!]
\centering
\includegraphics[width=0.7\textwidth]{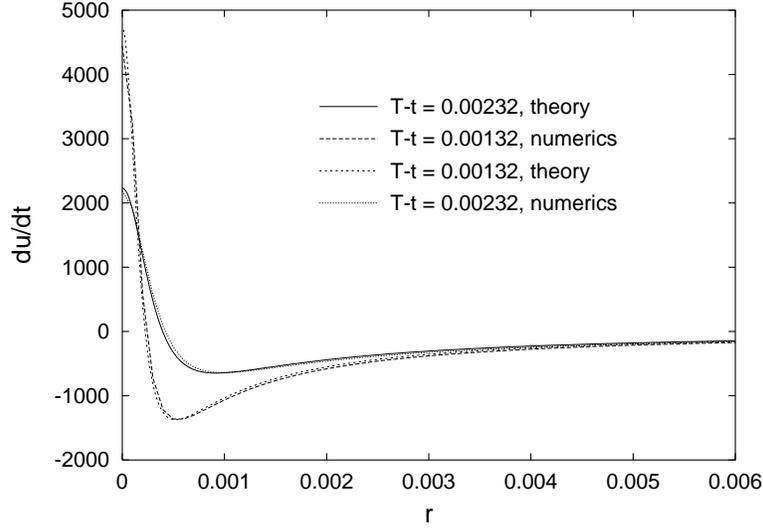}
\vskip 0.10cm
 \caption{\small{
For $p=7$ the time derivative $u_t(t,r)$ of marginally critical
solutions is plotted
 for two moments of time during the transient phase of evolution
 and compared to the  theoretical
 prediction $\partial u_1/\partial t = (T-t)^{-4/3} (U_1/3+\rho U'_1)$.
 The parameter $T$ is the same for both curves.}}
\end{figure}
\begin{figure}[h!]
\centering
\includegraphics[width=0.7\textwidth]{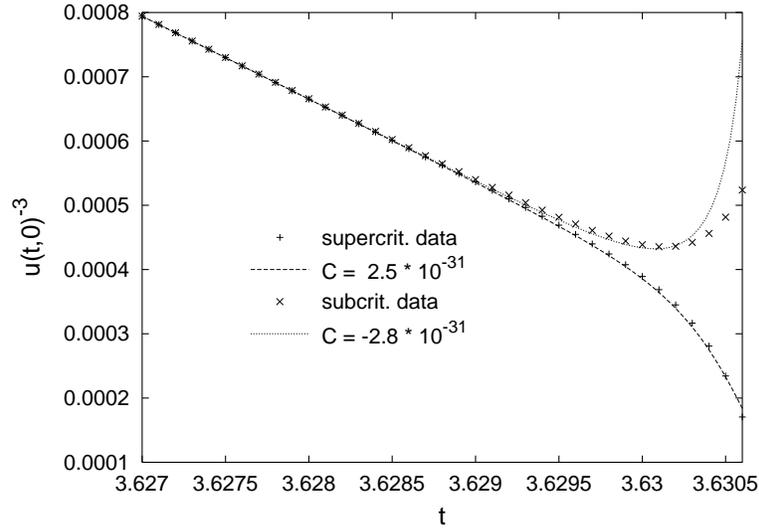}
\vskip 0.10cm
 \caption{\small{We  plot $u^{-3}(t,0)$ for the pair of marginally
 critical solutions corresponding to initial data (\ref{gauss4})
 with $A=A^*\pm 10^{-31}$. Initially these solutions are
 indistinguishable but eventually they split and depart from the
 critical solution towards blowup and dispersal, respectively. The
 theoretical curves, corresponding to equation (\ref{inter}) for
 $r=0$, with two different fitted coefficients $C$ are superimposed.}}
\end{figure}

 According to the picture of critical behaviour described above,
the marginally critical solutions have the following form in the
intermediate asymptotics
  \begin{equation}\label{inter}
  u(t,r) = (T-t)^{-\frac{1}{3}} U_1(\rho) + C(A) (T-t)^{-\lambda_1-\frac{1}{3}} \xi_1(\rho) + \mbox{damped modes} ,
  \end{equation}
  where $\xi_1$ is the single unstable mode about $u_1$ with the eigenvalue
  $\lambda_1=11.6442$.
  A small constant $C(A)$, which is the only vestige
   of  initial data, quantifies an admixture of the unstable mode -- for
   precisely critical data $C(A^*)=0$.  We show in figure 9 that the
   departure from the critical solution proceeds in agreement with
   equation (\ref{inter}).
 \noindent\subsection*{$p=5:$} We know from section
2 that in this case there are no nontrivial self-similar
solutions. However, since for $p=5$ the energy is scale invariant,
static solutions with finite energy are possible. Indeed, it is
well known that equation (\ref{eqspher}) has the finite energy
solution
\begin{equation}\label{sol5_stat}
 u_S(r) = (1 +\frac{1}{3} r^2)^{-\frac{1}{2}}.
\end{equation}
Rescalings of this solution generate the orbit of static solutions
$u_S^L=L^{-1/2} u_S(r/L)$. To determine the linear stability of
this solution we plug $u(t,r)=u_S(r) + e^{ikt} v(r)$ into
(\ref{eqspher}) and linearize. We get the eigenvalue problem in
the form of the radial Sch\"odinger equation
\begin{equation}\label{st_eigen}
-  v'' - \frac{2}{r} v' + V v = k^2 v, \quad
V=-\frac{5}{(1+\frac{1}{3} r^2)^2}.
\end{equation}
Notice that the perturbation induced by rescaling
\begin{equation}\label{zeromode}
    v_0(r)= -\frac{d}{dL} u_S^L(r)\Bigr\rvert_{L=1}
    =\frac{\frac{1}{2}-\frac{r^2}{6}}{(1+\frac{1}{3} r^2)^{\frac{3}{2}}}.
\end{equation}
satisfies equation (\ref{st_eigen}) for $k^2$. This is so  called
zero mode. Since the zero mode has one node, it follows by the
standard result from Sturm-Liouville theory that the potential $V$
has exactly  one bound state, $k_1^2<0$, which means that there is
exactly one growing mode $e^{\lambda_1 t} v_1(r)$, where
$\lambda_1=\sqrt{-k_1^2}$.
 Numerical calculation gives  $\lambda_1 \approx 1.1$.

  Thus, according to our preceding discussion,
 the solution $u_S$ is a
 candidate for a critical solution.
 To verify this,  Szpak~\cite{szpak} has investigated the
  nonlinear evolution of the growing mode. For initial data of the form
  $u(0,r)=u_S(r)+\epsilon v_1(r)$, $u_t(0,r)=\epsilon \lambda_1 v_1(r)$, he found that, depending
   on the
   sign of the amplitude $\epsilon$, the solution either disperses or blows up in
   finite time. This confirmed the expectation that in fact $u_S$ is the critical
   solution sitting on the saddle separating blowup from
   dispersal. Applying bisection to the family of initial data
   (\ref{gauss4}) we have obtained the solution $u_S$ as the
   intermediate attractor with pretty long lifetime.
We refer the reader to  \cite{szpak} for more
 details, in particular the analysis of convergence to $u_S$.
\vspace{-0.3cm}
\subsection*{$p=3:$} In this case we were not able to identify a
critical solution because of two reasons. First, in contrast to
the cases described above, in $p=3$ we do not have a good
candidate for the critical solution. The only potential candidate
 is the self-similar solution $u_1$ with one unstable mode,
however, as mentioned in section 2, this solution is singular
outside the past light cone of the blowup point and therefore
cannot be a \emph{bona fide} critical solution. Second, we face
the following difficulty when trying to determine $A^*$. As we
approach the expected value of $A^*$ from above the wave initially
shrinks but at some later time $t_1$ it bounces back and expands
outside with decreasing amplitude. During this period of evolution
the amplitude of outer wave front decreases faster then the
amplitude of the solution at the center, so a flat, slowly
decreasing with time central region develops. After some time
$t_2$ this central part of solution returns and starts growing
again to form a singularity at $r=0$. If we decrease $A$ further,
the time of bounce $t_1$ almost does not change but the return
time $t_2$ increases significantly. Therefore, approaching $A^*$
we have to evolve the solution longer and longer on larger and
larger grids. Since the numerical grid is always finite, we cannot
tell if an expanding wave which leaves the grid represents a
genuine dispersion or a singular solution with large return time
$t_2$. Figures 10 and 11 illustrate this difficulty.
\begin{figure}[h!]
\centering
\includegraphics[width=0.63\textwidth]{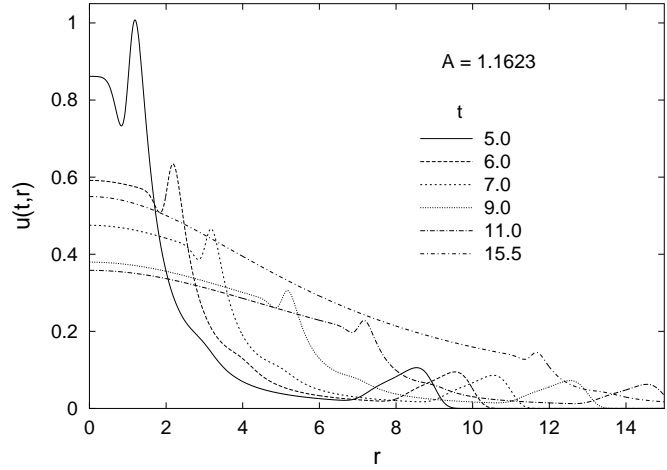}
\vskip 0.15cm
 \caption{\small{For $p=3$ we plot the snapshots from the evolution of the wave
that has bounced back from the center. After the bounce the
amplitude at the center initially decreases but later the wave
returns and the amplitude starts growing again.}}
\end{figure}
\begin{figure}[h!]
\centering
\includegraphics[width=0.63\textwidth]{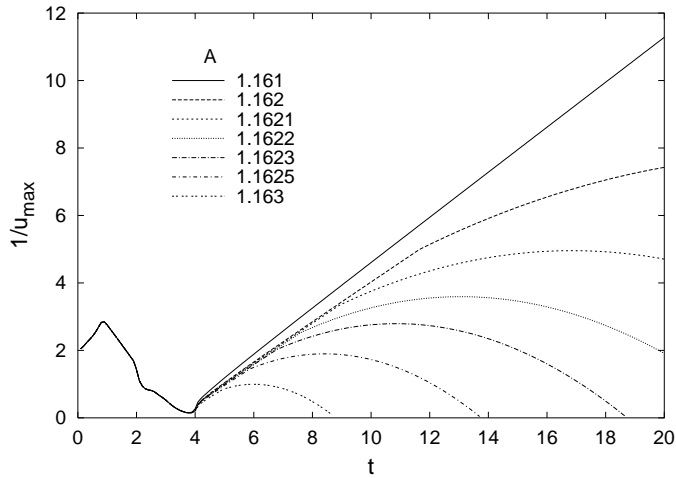}
\vskip 0.15cm
 \caption{\small{The same data as in figure 10. The first minimum of $1/u_{max}$ corresponds to the bounce.
 The second local maximum corresponds to the return.}}
\end{figure}
\section{Conclusions}
We have studied formation of singularities for the spherically
symmetric semilinear wave equation with the focusing power
nonlinearity $u^p$ for three representative values of the exponent
$p$: $p=3$ (subcritical case), $p=5$ (critical case), and $p=7$
(supercritical case). We showed that in all these cases the
asymptotic behaviour of blowup can be understood in terms of
decaying perturbations about the fundamental (homogeneous in
space) self-similar solution. We showed also that the nature of
the critical solution, whose codimension-one stable manifold
separates blowup form dispersal, depends on $p$: for $p=7$ the
critical solution is self-similar while  for $p=5$ it is static.
For $p=3$ we were not able to identify a critical solution --
determining the character of the threshold for blowup in this case
remains an open problem.
\section*{Acknowledgment}
 This research  was supported in part
by the KBN grant 2 P03B 006 23.

\end{document}